\documentclass{ws-procs975x65}

\begin{document}

\markboth{Francisco S. N. Lobo}{Traversable wormholes supported by
cosmic accelerated expanding equations of state}

\wstoc{Traversable wormholes supported by cosmic accelerated
expanding equations of state}{Francisco S. N. Lobo}

\title{TRAVERSABLE WORMHOLES SUPPORTED BY
COSMIC ACCELERATED EXPANDING EQUATIONS OF STATE}

\author{FRANCISCO S. N. LOBO}

\address{Centro de Astronomia
e Astrof\'{\i}sica da Universidade de Lisboa,\\
Campo Grande, Ed. C8 1749-016 Lisboa, Portugal\\
\email{flobo@cosmo.fis.fc.ul.pt}}

\begin{abstract}

We explore the possibility that traversable wormholes be supported
by specific equations of state responsible for the present
accelerated expansion of the Universe, namely, phantom energy, the
generalized Chaplygin gas, and the van der Waals quintessence
equation of state.

\end{abstract}

\keywords{Traversable wormholes; dark energy.}

\bodymatter

\section*{}

We shall explore the possibility that traversable wormholes
\cite{Morris} be supported by specific equations of state
responsible for the late time accelerated expansion of the
Universe, namely, phantom energy, the generalized Chaplygin gas,
and the van der Waals quintessence equation of state. Firstly,
phantom energy possesses an equation of state of the form
$\omega\equiv p/\rho<-1$, consequently violating the null energy
condition (NEC), which is a fundamental ingredient necessary to
sustain traversable wormholes. Thus, this cosmic fluid presents us
with a natural scenario for the existence of wormhole
geometries~\cite{phantomWH,Sushkov,dynphantomWH}. Secondly, the
generalized Chaplygin gas (GCG) is a candidate for the unification
of dark energy and dark matter, and is parametrized by an exotic
equation of state given by $p_{ch}=-A/\rho_{ch}^{\alpha}$, where
$A$ is a positive constant and $0<\alpha \leq 1$. Within the
framework of a flat Friedmann-Robertson-Walker cosmology the
energy conservation equation yields the following evolution of the
energy density
$\rho_{ch}=\left[A+Ba^{-3(1+\alpha)}\right]^{1/(1+\alpha)}$, where
$a$ is the scale factor, and $B$ is normally considered to be a
positive integration constant to ensure the dominant energy
condition (DEC). However, it is also possible to consider $B<0$,
consequently violating the DEC, and the energy density is an
increasing function of the scale function \cite{Lopez-Madrid}. It
is in the latter context that we shall explore exact solutions of
traversable wormholes supported by the GCG \cite{chaplyginWH}.
Thirdly, the van der Waals (VDW) quintessence equation of state,
$p=\gamma \rho/(1-\beta\rho)-\alpha \rho^2$, is an interesting
scenario for describing the late universe, and seems to provide a
solution to the puzzle of dark energy, without the presence of
exotic fluids or modifications of the Friedmann equations. Note
that $\alpha,\beta \rightarrow 0$ and $\gamma <-1/3$ reduces to
the dark energy equation of state. The existence of traversable
wormholes supported by the VDW equation of state shall also be
explored \cite{VDW-wh}.
Despite of the fact that, in a cosmological context, these cosmic
fluids are considered homogeneous, inhomogeneities may arise
through gravitational instabilities, resulting in a nucleation of
the cosmic fluid due to the respective density perturbations.
Thus, the wormhole solutions considered in this work may possibly
originate from density fluctuations in the cosmological
background.

The spacetime metric representing a spherically symmetric and
static wormhole geometry is given by (with $c=G=1$)
\begin{equation}
ds^2=-e ^{2\Phi(r)}\,dt^2+\left[1- b(r)/r\right]^{-1}dr^2+r^2
\,(d\theta ^2+\sin ^2{\theta} \, d\phi ^2)
\label{metricwormhole}\,,
\end{equation}
where $\Phi(r)$ and $b(r)$ are arbitrary functions of the radial
coordinate, $r$ \cite{Morris}. The latter has a range that
increases from a minimum value at $r_0$, corresponding to the
wormhole throat, to infinity. One may also consider a cut-off of
the stress-energy tensor at a junction radius $a$. The fundamental
properties of traversable wormhole are\cite{Morris}: The flaring
out condition of the throat, given by $(b-b'r)/b^2>0$, which
reduces to $b'(r_0)<1$ at the throat $b(r_0)=r_0$; the condition
$1-b(r)/r>0$, i.e., $b(r)<r$, is imposed; and the absence of event
horizons, which are identified as the surfaces with
$e^{2\Phi}\rightarrow 0$, so that $\Phi(r)$ must be finite
everywhere.

Using the Einstein field equation, $G_{\mu\nu}=8\pi \,T_{\mu\nu}$,
we obtain the relationships
\begin{eqnarray}
b'&=&8\pi r^2 \rho  \,,  \qquad \Phi'=\frac{b+8\pi r^3
p_r}{2r^2(1-b/r)}  \,,   \qquad p_r'=\frac{2}{r}\,(p_t-p_r)-(\rho
+p_r)\,\Phi ' \label{rhoWH}\,,
\end{eqnarray}
where $'=d/dr$. $\rho(r)$ is the energy density, $p_r(r)$ is the
radial pressure, and $p_t(r)$ is the tangential pressure. The
strategy we shall adopt is to impose an equation of state,
$p_r=p_r(\rho)$, which provides four equations, together with Eqs.
(\ref{rhoWH}). However, we have five unknown functions of $r$,
i.e., $\rho(r)$, $p_r(r)$, $p_t(r)$, $b(r)$ and $\Phi(r)$.
Therefore, to fully determine the system we impose restricted
choices for $b(r)$ or
$\Phi(r)$\cite{phantomWH,chaplyginWH,VDW-wh}. It is also possible
to consider plausible stress-energy components, and through the
field equations determine the metric fields\cite{Sushkov}.

Now, using the equation of state representing phantom energy,
$p_r=\omega \rho$ with $\omega<-1$, and taking into account Eqs.
(\ref{rhoWH}), we have the following condition
\begin{equation}
\Phi'(r)=\frac{b+\omega rb'}{2r^2\left(1-b/r \right)} \,.
            \label{EOScondition}
\end{equation}
For instance, consider a constant $\Phi(r)$, so that Eq.
(\ref{EOScondition}) provides $b(r)=r_0(r/r_0)^{-1/\omega}$, which
corresponds to an asymptotically flat wormhole geometry. It was
shown that this solution can be constructed, in principle, with
arbitrarily small quantities of averaged null energy condition
violating phantom energy, and the traversability conditions were
explored\cite{phantomWH}. The dynamic stability of these phantom
wormholes were also analyzed\cite{dynphantomWH}, and we refer the
reader to \cite{phantomWH,Sushkov} for further examples.

Relative to the GCG gas equation of state, $p_r=-A/\rho^{\alpha}$,
using Eqs. (\ref{rhoWH}), we have the following condition
\begin{equation}
2r\left(1-\frac{b}{r} \right)\Phi'(r)=-Ab'\left(\frac{8\pi r^2
}{b'}\right)^{1+\alpha}+\frac{b}{r} \,.
            \label{GCGEOScondition}
\end{equation}
Solutions of the metric (\ref{metricwormhole}), satisfying Eq.
(\ref{GCGEOScondition}) are denoted ``Chaplygin wormholes''. To be
a generic solution of a wormhole, the GCG equation of state
imposes the following restriction $A<(8\pi r_0^2)^{-(1+\alpha)}$,
consequently violating the NEC. However, for the GCG cosmological
models it is generally assumed that the NEC is satisfied, which
implies $\rho \geq A^{1/(1+\alpha)}$. The NEC violation is a
fundamental ingredient in wormhole physics, and it is in this
context that the construction of traversable wormholes, i.e., for
$\rho < A^{1/(1+\alpha)}$, are explored. Note that as emphasized
in \cite{Lopez-Madrid}$\,$, considering $B<0$ in the evolution of
the energy density, one also deduces that $\rho_{ch} <
A^{1/(1+\alpha)}$, which violates the DEC. We refer the reader
to$\,$\cite{chaplyginWH} for specific examples of Chaplygin
wormholes, where the physical properties and characteristics of
these geometries were analyzed in detail. The solutions found are
not asymptotically flat, and the spatial distribution of the
exotic GCG is restricted to the throat vicinity, so that the
dimensions of these Chaplygin wormholes are not arbitrarily large.

Finally, consider the VDW equation of state for an inhomogeneous
spherically symmetric spacetime, given by $p_r=\gamma
\rho/(1-\beta \rho)-\alpha \rho^2$. Equations (\ref{rhoWH})
provide the following relationship
\begin{equation}
2r\left(1-\frac{b}{r}\right) \Phi'=\frac{b}{r}+\frac{\gamma
b'}{1-\frac{\beta b'}{8\pi r^2}} -\frac{\alpha b'^2}{8\pi r^2} \,.
            \label{vdWEOS}
\end{equation}
It was shown that traversable wormhole solutions may be
constructed using the VDW equation of state, which are either
asymptotically flat or possess finite dimensions, where the exotic
matter is confined to the throat neighborhood \cite{VDW-wh}. The
latter solutions are constructed by matching an interior wormhole
geometry to an exterior vacuum Schwarzschild vacuum, and we refer
the reader to$\,$\cite{VDW-wh} for further details.


In concluding, it is noteworthy the relative ease with which one
may theoretically construct traversable wormholes with the exotic
fluid equations of state used in cosmology to explain the present
accelerated expansion of the Universe. These traversable wormhole
variations have far-reaching physical and cosmological
implications, namely, apart from being used for interstellar
shortcuts, an absurdly advanced civilization may convert them into
time-machines, probably implying the violation of causality.


\begin{thebibliography}{99}


\bibitem{Morris}
M. S. Morris and K. S. Thorne, Am. J. Phys. {\bf 56}, 395 (1988).

\bibitem{phantomWH}
F. S. N. Lobo, Phys. Rev. D {\bf 71}, 084011 (2005).

\bibitem{Sushkov}
S.~Sushkov, Phys. Rev. D {\bf 71}, 043520 (2005).

\bibitem{dynphantomWH}
F. S. N. Lobo, Phys. Rev. D {\bf 71}, 124022 (2005).


\bibitem{Lopez-Madrid}
M. Bouhmadi-L\'{o}pez and J. A. J. Madrid, JCAP {\bf 0505}, 005
(2005).

\bibitem{chaplyginWH}
F. S. N. Lobo, Phys. Rev. D {\bf 73} 064028 (2006).


\bibitem{VDW-wh}
F. S. N. Lobo, ``Van der Waals quintessence stars,''
[arXiv:gr-qc/0610118].

\end{thebibliography}
\end{document}